\RequirePackage{fix-cm}
\documentclass[smallextended]{svjour3}       
\smartqed  
\usepackage{graphicx}

\usepackage{url}
\usepackage{color}
\journalname{Multimedia Tools and Applications}

\begin{document}

\title{Open Annotations on Multimedia Web Resources}

\author{Bernhard Haslhofer\and
        Robert Sanderson\and
        Rainer Simon\and
        Herbert van de Sompel
}

\institute{Bernhard Haslhofer \at
              Cornell University,
              Department of Information Science \\
              301 College Avenue, Ithaca, NY 14850, USA \\
              \email{bernhard.haslhofer@cornell.edu}           
           \and
           Robert Sanderson and Herbert van de Sompel \at
              Los Alamos National Laboratory \\
              Los Alamos, NM 87544, USA\\
              \email{rsanderson,herbertv@lanl.gov}
           \and
           Rainer Simon \at
           Austrian Institute of Technology\\
           Donau-City-Str. 1, A-1220 Vienna, Austria \\
           \email{rainer.simon@ait.ac.at}
}

\date{Received: \today / Accepted: date}

\maketitle

\begin{abstract}

Many Web portals allow users to associate additional information with
existing multimedia resources such as images, audio, and video. However,
these portals are usually closed systems and user-generated annotations
are almost always kept locked up and remain inaccessible to the Web of Data.
We believe that an important step to take is the integration of multimedia
annotations and the Linked Data principles.
We present the current state of the Open Annotation Model, explain our design
rationale, and describe how the model can represent user annotations
on multimedia Web resources. Applying this model in Web portals and devices,
which support user annotations, should allow clients
to easily publish and consume, thus exchange annotations on multimedia Web
resources via common Web standards.


\end{abstract}


\section{Introduction}\label{sec:intro}

\begin{figure}
  \centering
  \includegraphics[width=0.75\textwidth]{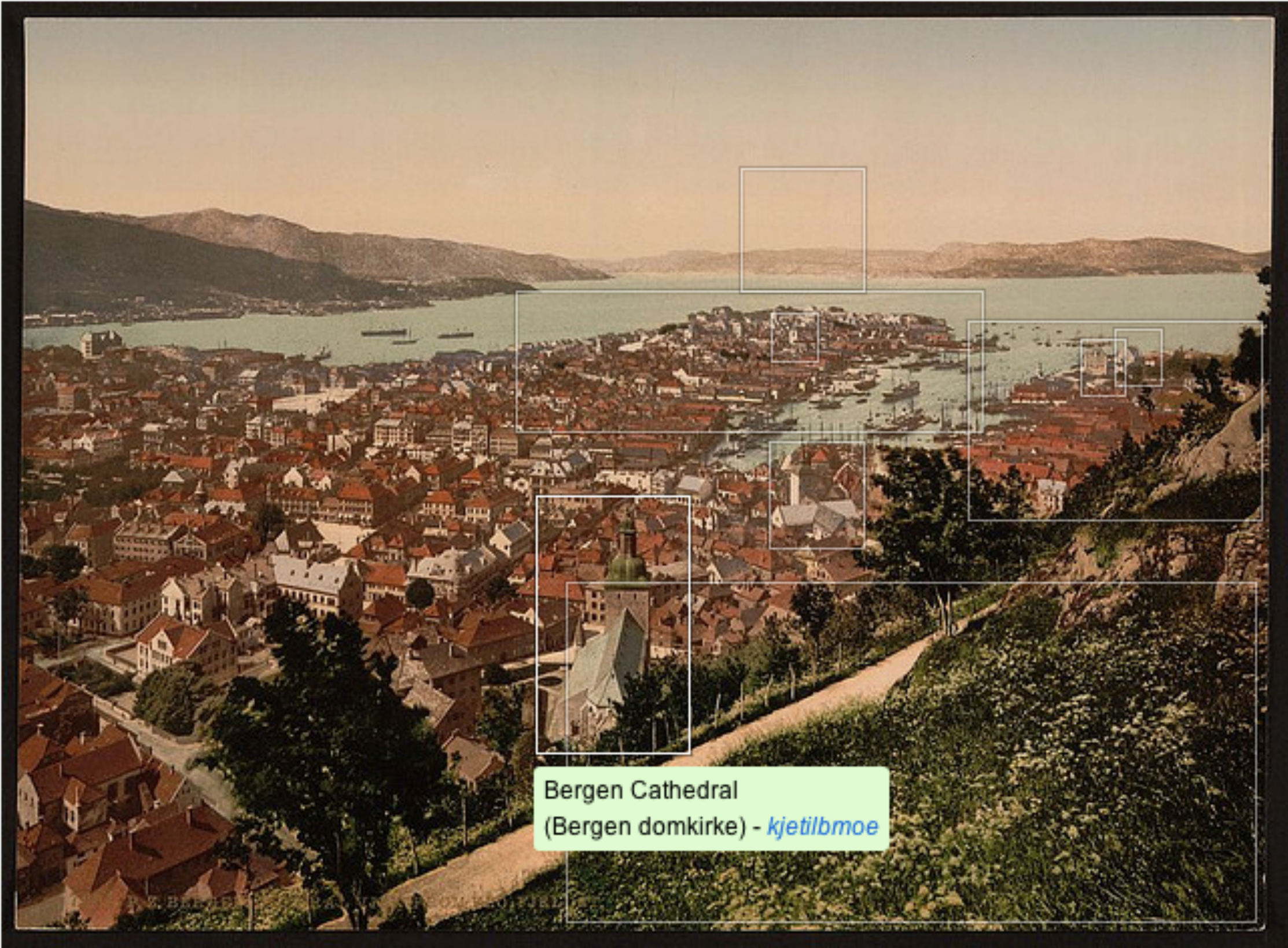}
  \caption{An annotation example on Flickr.}
  \label{fig:flickr_example}
\end{figure}

Youtube and Flickr are examples of large-scale Web portals that allow users to annotate multimedia resources by adding textual notes and comments to images or videos. Figure~\ref{fig:flickr_example} shows an image\footnote{\url{http://www.flickr.com/photos/library_of_congress/3175009412/}} from the Flickr Commons collection contributed by the Library of Congress. Flickr users added several annotations to specific image segments, one of them telling us that this picture shows a cathedral in Bergen.

Annotations describe resources with additional information, which is valuable to other users, who are searching and browsing resource collections. They are also important for underlying information systems, which can exploit the high-level descriptive semantics of annotations in combination with automatically extracted low-level features, such as image size and color, to implement search and retrieval over multimedia resources. Taking the previous example, users who are searching for ``Bergen'' or even ``Bergen Cathedral'' will now find this particular image in Flickr, because some user provided this descriptive information in textual form.

Annotations are also becoming an increasingly important component in scholarly cyber-infrastructures (cf.~\cite{Bradley:2008kx}), which are often realized as Web systems. Therefore, a Web-based annotation model should fulfill several requirements. In the age of video blogging
and real-time sharing of geo-located images, the notion of solely textual annotations has become obsolete. Instead, \emph{multimedia} Web resources should be annotatable and also be able to be annotated onto other resources. Users often discuss multiple
segments of a resource, or multiple resources, in a single annotation
and thus the model should support multiple targets. An annotation framework should also follow the Linked Open Data guidelines~\cite{Heath:2011uq} to promote annotation sharing between systems. In order to avoid inaccurate or incorrect annotations, it must take the ephemeral nature of Web resources into account.

Annotations on the Web have many facets: a simple example could be a textual note or a tag (cf.,~\cite{Hunter2009}) annotating an image or video. Things become more complex when a particular paragraph in an HTML document annotates a segment (cf.,~\cite{Hausenblas:LDOW09}) in an online video or when someone draws polygon shapes on tiled high-resolution image sets. If we further extend the annotation concept, we could easily regard a large portion of Twitter tweets as annotations on Web resources. Therefore, in a generic and Web-centric conception, we regard an annotation as an association created between one \emph{body} resource and other \emph{target} resources, where the body must be somehow \emph{about} the target.

Annotea~\cite{Kahan:2001vn} already defines a specification for publishing annotations on the Web but has several shortcomings: (i) it was designed for the annotation of Web pages and provides only limited means to address segments in multimedia objects, (ii) if clients want to access annotations they need to be aware of the Annotea-specific protocol, and (iii) Annotea annotations do not take into account that Web resources are very likely to have different states over time.

Throughout the years several Annotea extensions have been developed to deal with these and other shortcomings: Koivunnen~\cite{Koivunen:2006s} introduced additional types of annotations, such as \emph{bookmark} and \emph{topic}. Schroeter and Hunter~\cite{Schroeter:uq} proposed to express segments in media-objects by using \emph{context} resources in combination with formalized or standardized descriptions to represent the context, such as SVG or complex datatypes taken from the MPEG-7 standard. Based on that work, Haslhofer et al.~\cite{Haslhofer:2009ve} introduce the notion of \emph{annotation profiles} as containers for content- and annotation-type specific Annotea extensions and suggested that annotations should be dereferencable resources on the Web, which follow the Linked Data guidelines. However, these extensions were developed separately from each other and inherit some of the above-mentioned Annotea shortcomings.

In this article we describe the \emph{Open Annotation Model}\footnote{Open Annotation Model: Beta Data Model Guide \url{http://www.openannotation.org/spec/beta/}}, which is currently being developed in an international collaboration. It applies a Web- and resource-centric view on annotations and defines a modular architecture, which has a simple base line model in its core. The model also provides means to address segments in multimedia resources either by encoding segment information using the Media Fragment URI specification or by introducing custom segment constraints for more complex annotation use cases. By allowing fixity and timestamp information on the resources involved in an annotation, it also takes into account the ephemeral nature of Web resources. Pulling together the functionalities provided by various, partly independent Annotea extensions is a major goal of this effort.

This article extends our work previously published in~\cite{Haslhofer:2011fk} by the following: it explains the design rationale that lead to the specification of the current annotation model and also the technical aspects of the model in more detail. It also contains an updated and extended related work section.


\section{Design Rationale}\label{sec:examples}

In this section we outline the design rationale that drives the specification of the Open Annotation Model. We give examples illustrating common requirements we found in several real-world annotation use cases (e.g., \cite{Sanderson:2011a,Verspoor:2005kx,Simon:2011vn}) and describe the reasons behind our design decisions. From each design decision we derived a set of guiding design principles, which are reflected in the current Open Annotation Model design.

\subsection{Annotations are qualified associations between resources}

\begin{figure}
  \centering
  \includegraphics[width=1\textwidth]{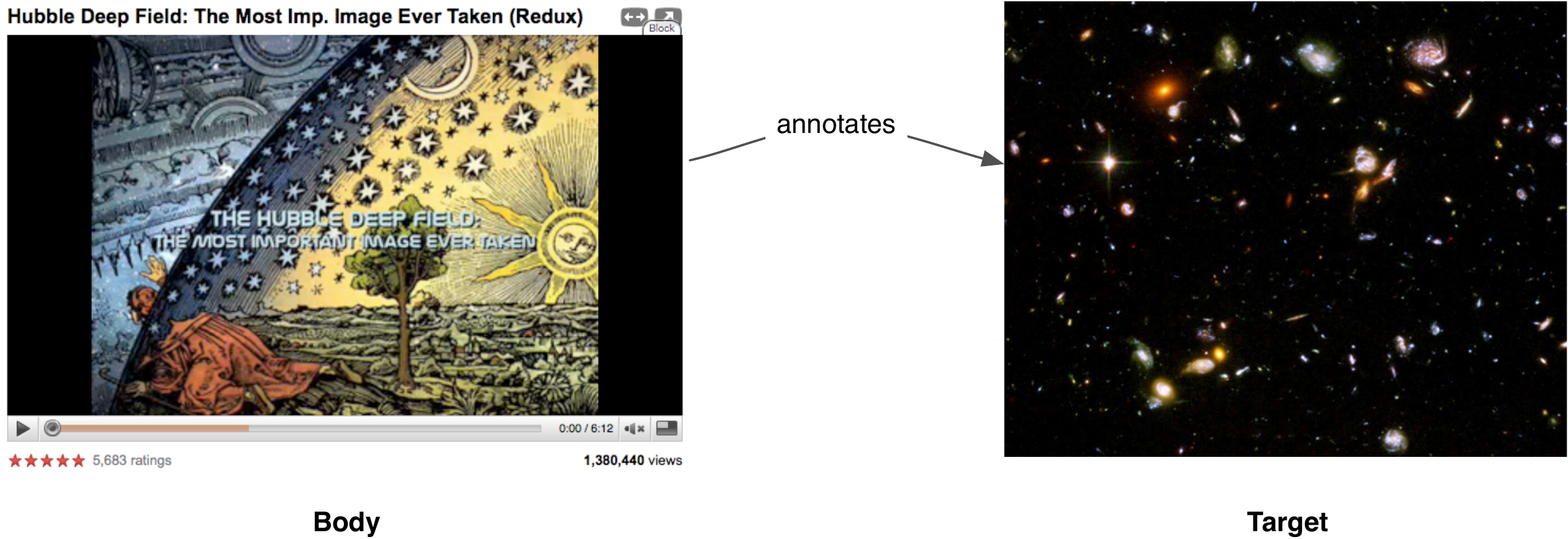}
  \caption{A Youtube video annotating an image on the Web.}
  \label{fig:hubble_example}
\end{figure}

Textual notes or tags on images, as in the Flickr example in Figure~\ref{fig:flickr_example}, occur frequently on the Web and are the simplest examples for Web annotations. In Open Annotation Model terms, these are \emph{annotations} that have a textual \emph{body} and an image resource as \emph{target}. However, in many scenarios the prevailing view that annotation bodies are textual is insufficient. Figure~\ref{fig:hubble_example} shows an annotation in which the body is not a textual note, but a video, which itself is an addressable Web resource identified by a URI. In order to cover such cases, we must abstract from purely textual annotation bodies and model them as resources that can be of any media type.

From a conceptual point of view, an annotation is an association between two resources, the \emph{body} and the \emph{target}. However, in most cases this association needs to be qualified and addressable in some way. Indicating the creator and creation date of a resource, or replying to existing annotations, are frequently occurring scenarios that require an annotation to be expressed as a first-class entity. Web annotations are then instances of an annotation and relate together the body and target resources that are involved in the annotation association. With this approach we follow a common design pattern, which is also known as Qualified Relation in Linked Data~\cite{Dodds:uq} or Association Class in UML.

In contrast to existing tagging models (see~\cite{KimEtAl2008}) the annotation \emph{creator} is not a mandatory core model entity. We believe that adopters of the Open Annotation Model already have user models in place or rely on open Web identities (e.g., a Google or Facebook account) and are most likely not willing to map their user models against another user model, which is specified as part of an annotation model. However, we encourage adopters to relate annotations with existing Web resources, which identify users.

From these considerations we derive the following guiding principles:

\begin{itemize}
    \item All core entities (annotation, body, target) must be resources.
    \item Annotations must allow for both body and target of any media type.
    \item Annotations, bodies, and targets can have different authorship.
\end{itemize}

\subsection{Annotations involve parts of multimedia resources}

Our introductory Flickr example illustrates how annotations can target specific segments in Web resources. In Figure~\ref{fig:hubble_example} we gave an example that involves multimedia resources: a video as body, and an image as target. There is strong dependency between these requirements because a resource's media type affects the way segments need to be addressed. Annotating an area in an image requires a different segment representation than those that target segments in the spatial and temporal dimensions of a video. Similarly, addressing text segments in a PDF document differs from addressing text in plain text or HTML documents. Since we abstract from purely textual annotation bodies, we must take into account that both the body and the target of an annotation can be resource segments. We could, for instance, refine our previous example by saying that some sequence within the video annotates a certain image segment.

The problem of addressing media segments, also known as fragment identification, is well known and will be explained in more detail in Section~\ref{subsec:rel_work_media_segments}. Since the Open Annotation Model will be implemented in Web environments and all resources involved in an annotation are Web resources identified by URIs, it should reuse the fragment identification mechanisms that are already defined as part of the Architecture of the World Wide Web~\cite{Jacobs2004} and extensions thereof, such as fragment construction rules for specific media types. However, many annotation use cases require more complex segment representations such as polygon regions in images, which cannot be expressed with available standards. Therefore, our guiding principles with respect to addressing segments in multimedia resources are:

\begin{itemize}
    \item Annotations must support resource segment addressing both on body and target resources.
    \item Preferably this should be done with (media) fragment URIs, but extensibility must be provided for cases in which the use of URIs for segment addressing is not possible.
\end{itemize}

\subsection{Annotation resources are ephemeral Web resources}

Previously we argued that all core entities of the Open Annotation Model must be first class Web resources, which are identified by URIs, if possible HTTP URIs. The great benefit of this approach is that existing technologies and solutions that can be applied for Web resources (e.g., mime-types, fragments, access control, etc.) also work for resources involved in a Web annotation without the necessity to include these aspects in an annotation specification. However, there is one big problem we inherit from the Web architecture and which is severe in the context of annotations: URI-addressable Web resources are ephemeral, which means that the representations obtained by dereferencing their URIs may change over time.

\begin{figure}
  \centering
  \includegraphics[width=1\textwidth]{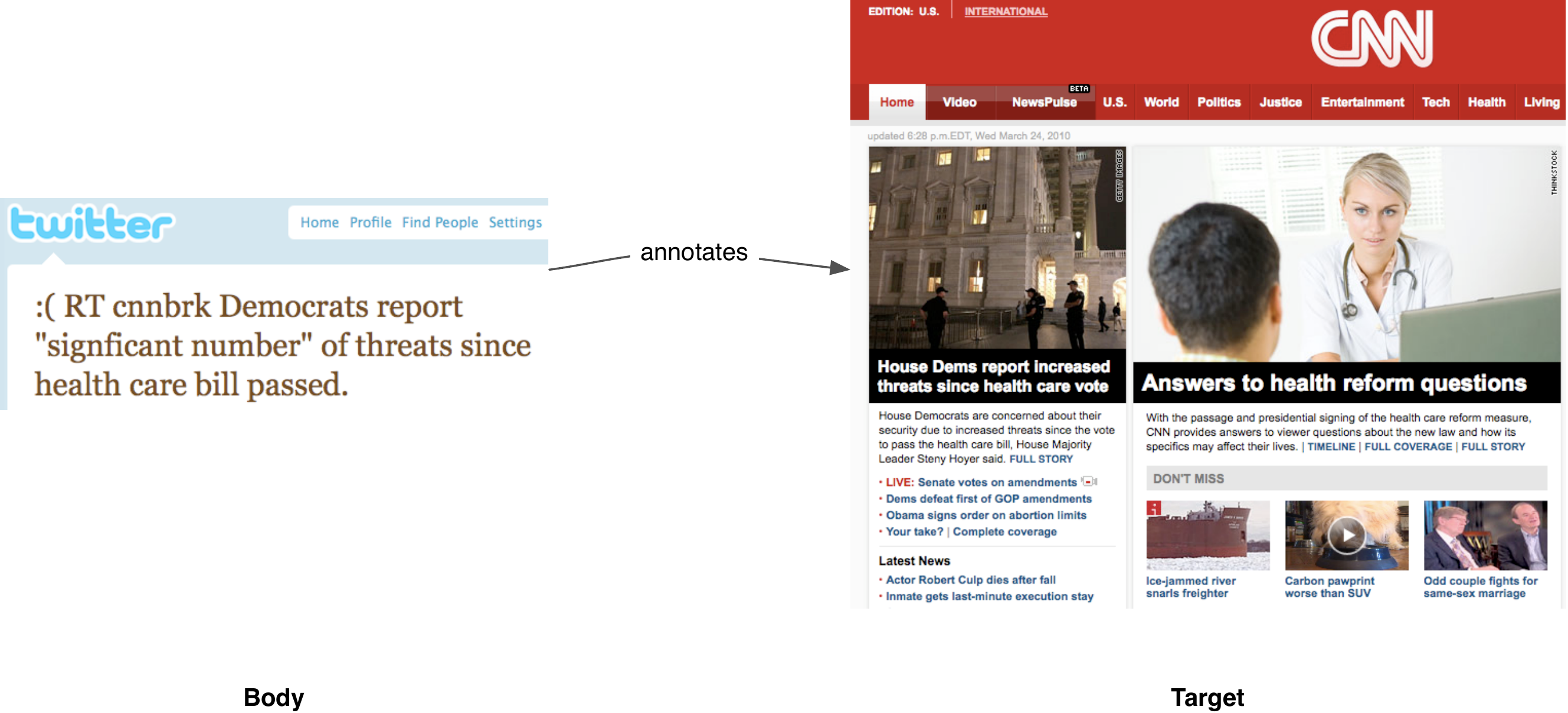}
  \caption{A Twitter tweet annotating the CNN website.}
  \label{fig:cnn_example}
\end{figure}

The annotation example in Figure~\ref{fig:cnn_example} shows a Twitter tweet annotating the CNN web site and illustrates the ephemerality problem: the tweet refers to the CNN page at a certain point in time and might be misinterpreted when the CNN main web site changes. Measures should be provided that can help in avoiding misinterpretations of annotations, including the expression of timestamps and fixity information for body and target resources.
 
\subsection{Annotations should be interoperable}

The focus of the Open Annotation Model is on sharing annotations on scholarly resources and therefore the model should have sufficient richness of expression to satisfy scholars' needs. However, in order to maximize the likelihood of adoption, the model should also be an interoperability framework readily applicable in other domains. One possibility to achieve this is to follow a modular and extensible modeling approach, with a generally applicable baseline model and domain- or media-type specific extensions. An annotation client should implement at least the baseline model and, if possible, provide fallback behavior for annotations that contain domain-specific extensions the client might not be aware of.

In order to increase the likelihood of adoption, and in alignment with the goal of sharing annotations, no client-server protocol for publishing, updating, or deleting annotations will be specified. Rather, the specifications will take a perspective whereby clients publish annotations to the Web and make them discoverable using common Web approaches. Such an approach does not require a preferred annotation server for a client, yet it does not preclude one either.

Our guiding principles to achieve annotation interoperability and widespread adoption are:
\begin{itemize}
    \item The Open Annotation Model should have a simple but expressive baseline model defining top-level classes/entities and properties/relationships.
    \item The baseline model should be extensible.
    \item Annotation protocols are out of scope.
\end{itemize}


\section{The Baseline Model}\label{sec:baseline_model}

The Open Annotation data model draws from various extensions of
Annotea to form a cohesive whole. The Web architecture and Linked Data
guidelines are foundational principles, resulting in a specification
that can be applied to annotate any set of Web resources. At the time
of this writing, the specification, which is available at
\url{http://www.openannotation.org/spec/beta/}, is still under
development. In the following, we describe the major technical
building blocks of the Open Annotation data model and use some simple examples
to illustrate how to apply the model in practice. For further
examples, which also cover different use cases, such as the annotation of medieval
manuscripts, we refer to the online documentation.

\begin{figure}
  \centering
  \includegraphics[width=0.5\textwidth]{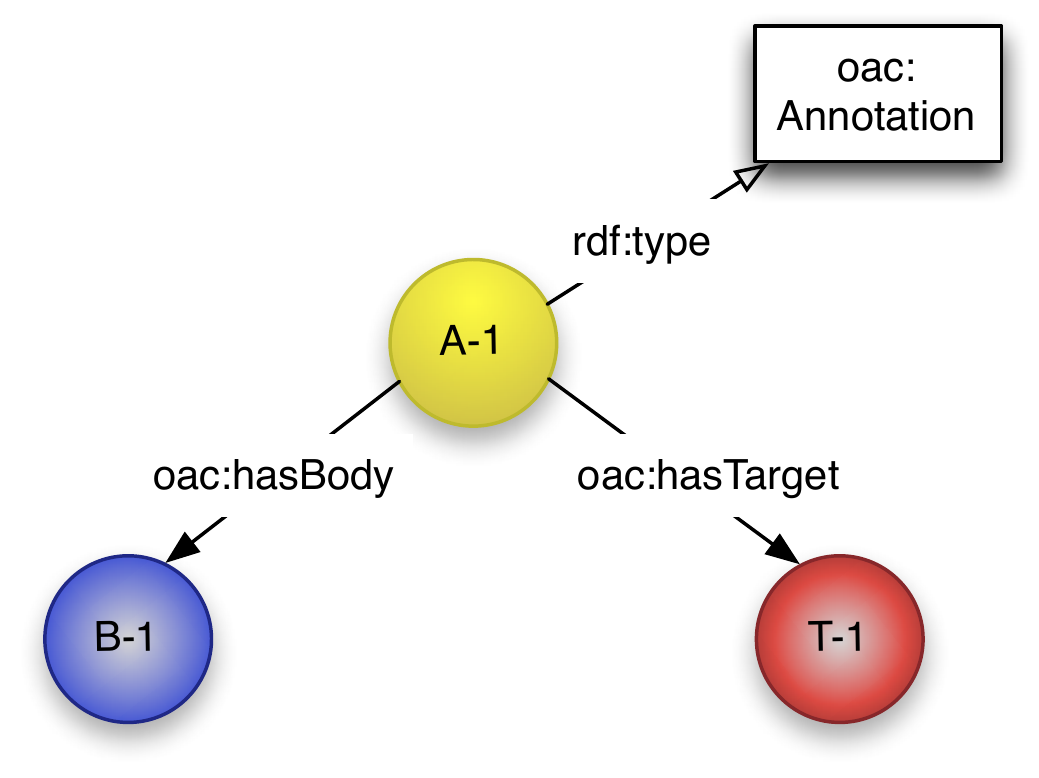}
  \caption{Open Annotation baseline data model.}
\label{fig:oac_data_model}
\end{figure}

Following its predecessors, the Open Annotation Model,
shown in Figure~\ref{fig:oac_data_model}, has three primary classes of
resources. In all cases below, the \texttt{oac} namespace prefix
expands to \url{http://www.openannotation.org/ns/}.

\begin{itemize}
	\item The \texttt{oac:Body} of the annotation (node B-1): This resource is the comment, metadata or other information that is created about another resource. The body can be any Web resource, of any media format, available at any URI. The model allows for either one body per annotation, or an annotation without any body, but not annotations with multiple bodies.
	
	\item The \texttt{oac:Target} of the annotation (node T-1): This is the resource that the body is about. Like the body, it can be any URI identified resource. The model allows for one or more targets per annotation.
	
	\item The \texttt{oac:Annotation} (node A-1): This resource is an RDF document, identified by an HTTP URI, that describes at least the body and target resources involved in the annotation as well as any additional properties and relationships (e.g., \texttt{dcterms:creator}). Dereferencing an annotation's HTTP URI returns a serialization in a permissible RDF format.
	
\end{itemize}

\begin{figure}
  \centering
  \includegraphics[width=0.75\textwidth]{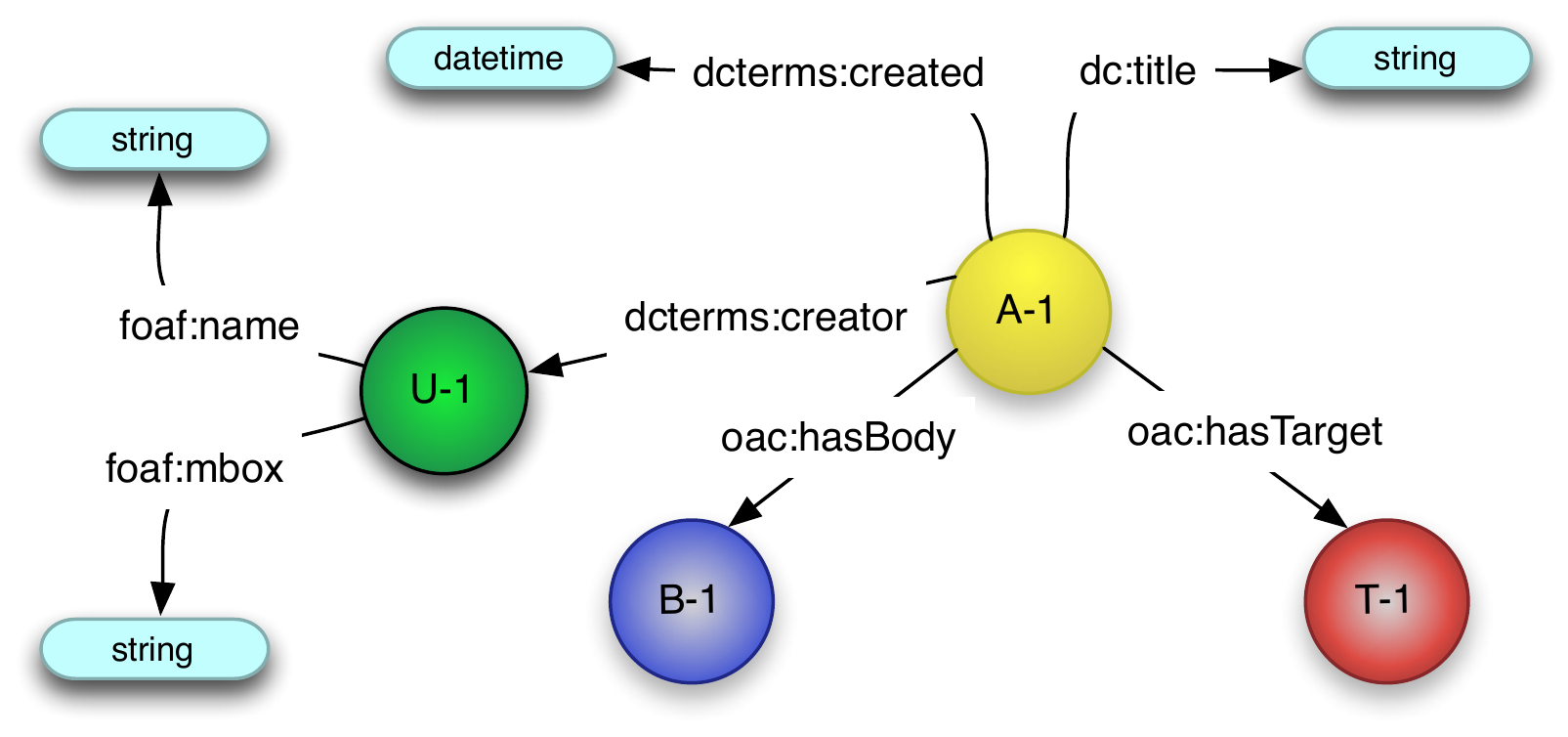}
  \caption{Additional properties and relationships in the Open Annotation baseline model.}
\label{fig:oac_data_model_add_prop}
\end{figure}

An annotation is a Web resource, which is identified by an HTTP URI and
returns an RDF document when being dereferenced. As with any RDF data, 
additional properties and relationships can be associated with any of the
resources. It is recommended that an annotation has a timestamp of when
the annotation relationship was created (\texttt{dcterms:created}) and a
reference to the agent that created it (\texttt{dcterms:creator}). Resources
referenced by additional relationships may themselves have additional
properties and relationships. Figure~\ref{fig:oac_data_model_add_prop} gives
an example of recommended and other possible (e.g., \texttt{dc:title}) properties
and relationships that can be added to the Open Annotation baseline model.
The set of properties and relationship in this example is by no means exhaustive.
Properties and relationships from other vocabularies may also be used. It is also
important to note that the creator and created timestamp of each of the three resource
types above may be different. An annotation might refer to
an annotation body created by a third party, perhaps from before the 
Open Annotation specification was published, and a target created by yet another party.

Similarly,
there may be additional subclasses of \texttt{oac:Annotation} that further specify
restrictions on the meaning of the annotation, such as a \emph{reply}; an
annotation where the single target is itself an annotation. This
allows chaining of annotations into a threaded discussion model.

If the body of an annotation is identified by a dereferencable HTTP
URI, as it is the case in Twitter, various blogging platforms, or
Google Docs, it can easily be referenced from an annotation. If a
client cannot create URIs for an annotation body, for instance because
it is an offline client, it can assign a unique non-resolvable URI
(called a URN) as the identifier for the body node. This approach can
still be reconciled with the Linked Data principles as servers that
publish such annotations can assign HTTP URIs they control to the
bodies, and express equivalence between the HTTP URI and the URN.

\begin{figure}
  \centering
  \includegraphics[width=0.75\textwidth]{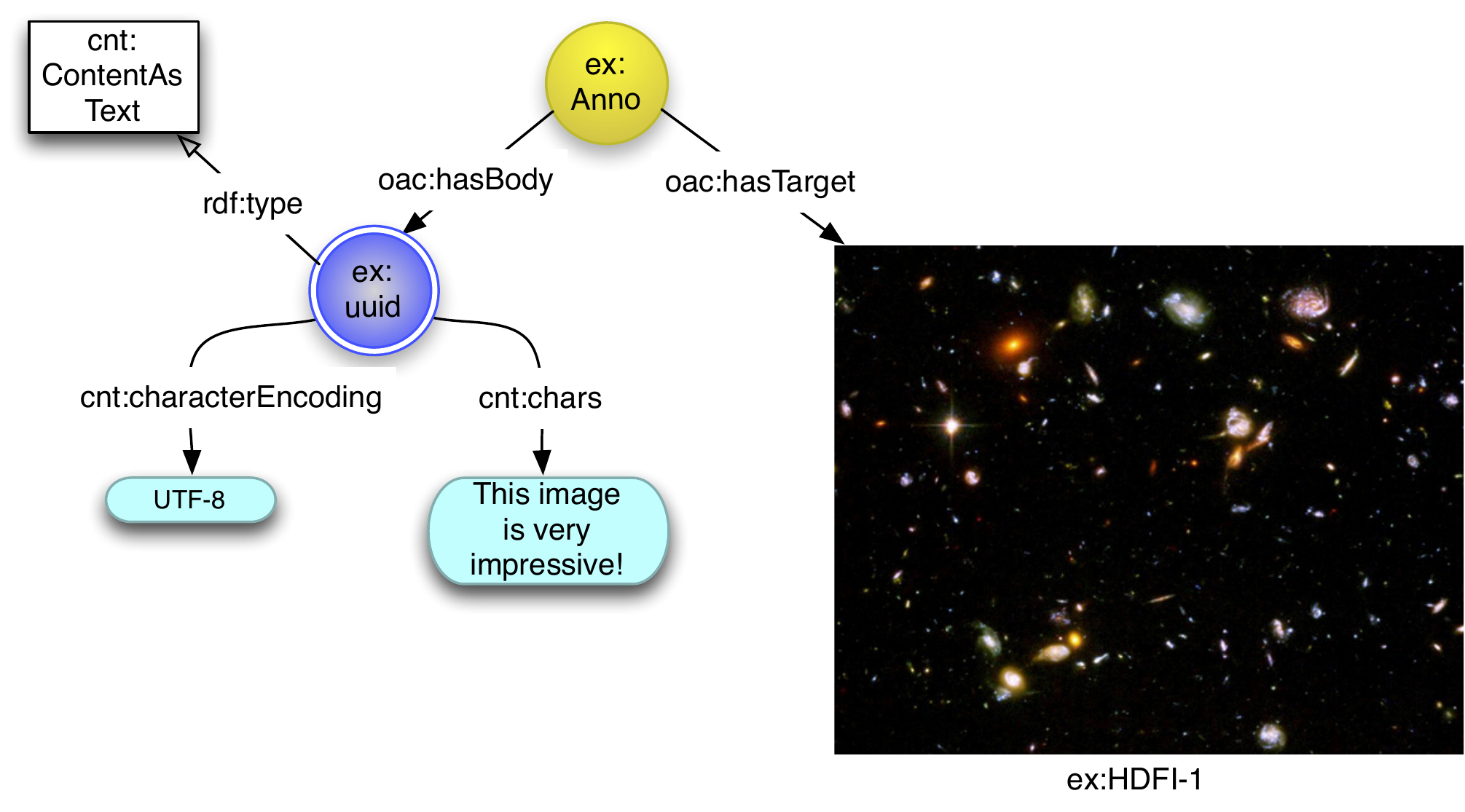}
  \caption{An example annotation with inline body.}
\label{fig:oac_data_model_inline_body_example}
\end{figure}

The Open Annotation Model allows the inclusion of information directly in the
annotation document by adding the representation of a resource inline
within the RDF document using the \emph{Content in RDF}
specification~\cite{Koch:2009uq} from the W3C. The example annotation in Figure~\ref{fig:oac_data_model_inline_body_example} shows how to express this:
the representation is the object of the \texttt{cnt:chars} predicate, and its character
encoding the object of the \texttt{cnt:characterEncoding} predicate.
Further classes from this specification include Base64 encoded
resources and XML encoded resources.


\section{Annotating Media Segments}\label{sec:media_segments}

Most of the use cases, which have been explored before specifying the model,
involved comments that were about a segment of a
resource, rather than the entire resource identified by a URI. The
data model allows two different methods of identifying and describing
the region of interest of a resource; either using a \emph{fragment URI}, or
a more expressive \emph{constraint} resource. It is clearly recommended to 
use fragment URIs whenever possible, because this method relies on normative
specifications, which brings interoperability with other applications. Only
if there are no appropriate URI fragment specifications available, the creators
should define their own constraints. 

\subsection{Describing segments with fragment URIs}

A fragment URI normally identifies a part of a resource, and the method 
for constructing and interpreting these URIs is dependent on the media
type of the resource. In general, fragment URIs are created by appending
a fragment that describes the section of interest to the URI of the full
resource, separated by a '\#' character (see \cite{Berners-Lee:2005uq}).

There are two main sources for existing fragment URI specifications, which
can both identify and describe how to discover a segment of interest within a
resource. The first is the set of Mime Type specification RFCs from the IETF.
This includes X/HTML (RFC 2854/3236), XML (RFC 3023), PDF (RFC 3778) and
Plain Text (RFC 5147). The second is W3C Media URIs specification~\cite{fragmentsURI:2011ab}, 
which is defined at a broader level to cover images, video and audio resources, 
regardless of the exact format. The following examples show how to apply these
specifications to define media-type specific URIs that describe a certain resource segment:

\begin{itemize}
    \item \url{http://www.example.net/foo.html#namedSection} identifies the section named as ``namedSection'' in an HTML document.
    \item \url{http://www.example.net/foo.pdf#page=10&viewrect=20,100,50,60} identifies a rectangle starting at 20 pixels in from the left, and 100 down from the top, with a width of 50 and a height of 60 in a PDF document.
    \item \url{http://www.example.net/foo.png#xywh=160,120,320,240} identifies a 320 by 240 box, starting at x=160 and y=120 in an image.
    \item \url{http://www.example.net/foo.mpg#t=npt:10,20 } identifies a sequence starting just before the 10th second, and ending just before the 20th in a video.
\end{itemize}

\begin{figure}
  \centering
  \includegraphics[width=1\textwidth]{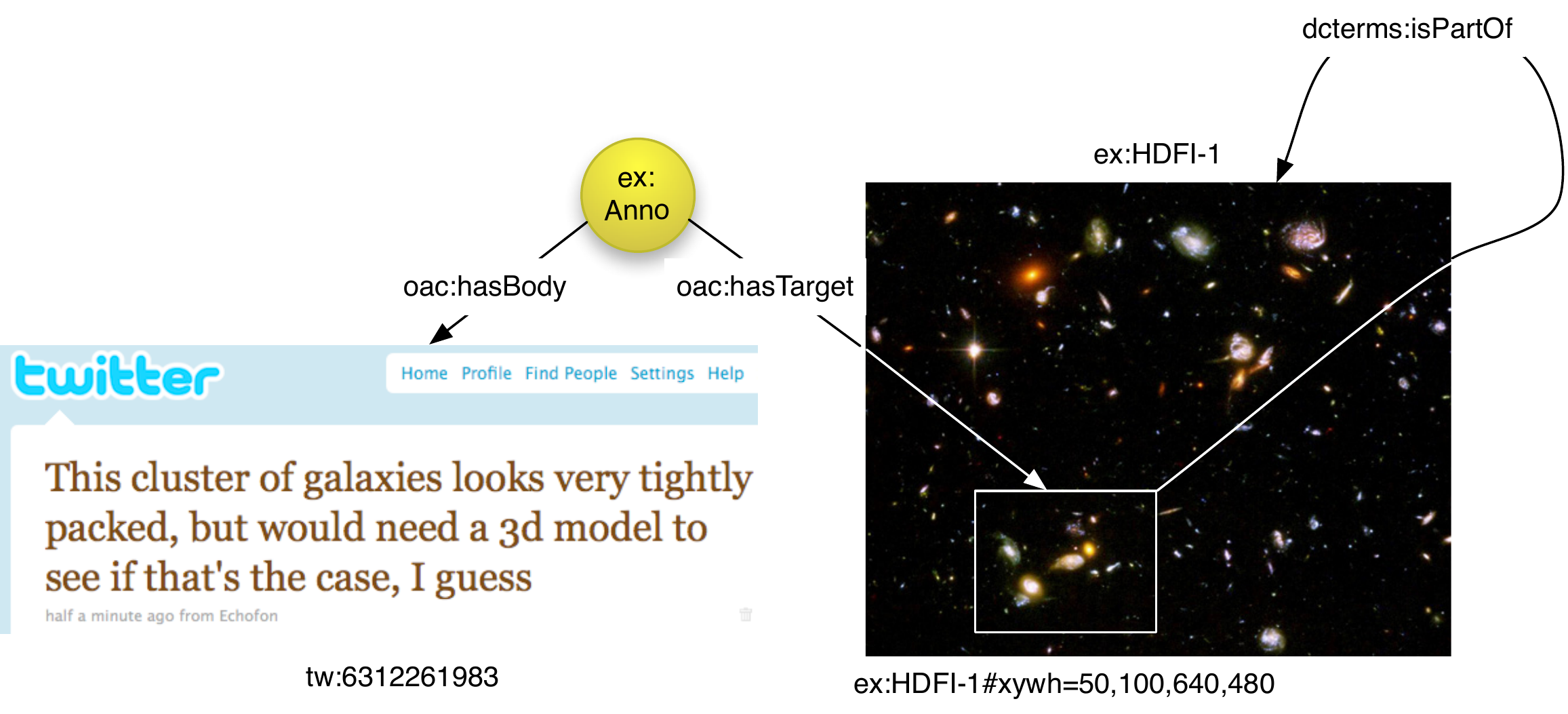}
  \caption{Annotating media segments using a fragment URI.}
\label{fig:media_segment_uri}
\end{figure}

We recommend that when a definition exists for how to construct a fragment URI for a particular document format, and such a fragment would accurately describe the section of interest for an annotation, then this technique should be used. It is recommended to also use \texttt{dcterms:isPartOf} with the full resource as the object, in order to make the annotation
more easily discoverable. Figure~\ref{fig:media_segment_uri} shows an example in which a tweet annotates a rectangular section in an image, which in turn is identified and described by a media fragment URI.

\subsection{Describing segments via constraint resources}

\begin{figure}
  \centering
  \includegraphics[width=0.75\textwidth]{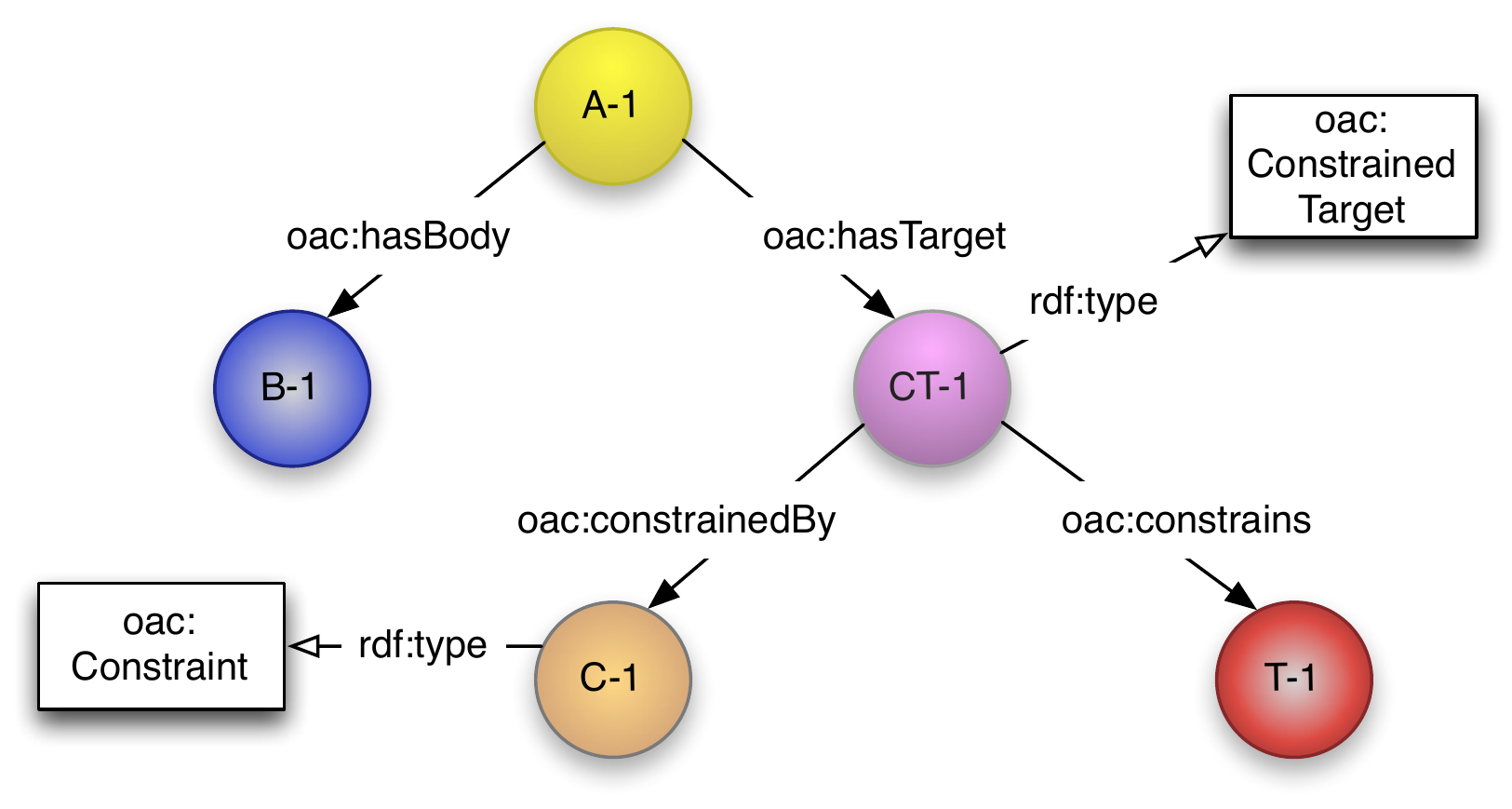}
  \caption{Constraint annotation targets in the Open Annotation Model.}
  \label{fig:oac_model_constraint_target}
\end{figure}

There are many situations when segments cannot be described with fragment 
URIs, but it is still desirable to be able to annotate a segment of a
resource.  For example, a non-rectangular section of an image, or a
segment of a resource with a format or media type that is not covered
by either fragment specification, such as a 3-dimensional model or a
dataset. To handle these situations, we introduce an \texttt{oac:Constraint}
resource that describes the segment of interest using an appropriate
standard, and a \texttt{oac:ConstrainedTarget} resource that identifies the segment
of interest. This \emph{constrained target} is the object of the \texttt{oac:hasTarget}
predicate of the \texttt{oac:Annotation}, and subsequently \texttt{oac:constrains}
the full target resource. Figure~\ref{fig:oac_model_constraint_target} shows
how constrained targets extend the Open Annotation baseline model.

\begin{figure}
  \centering
  \includegraphics[width=0.75\textwidth]{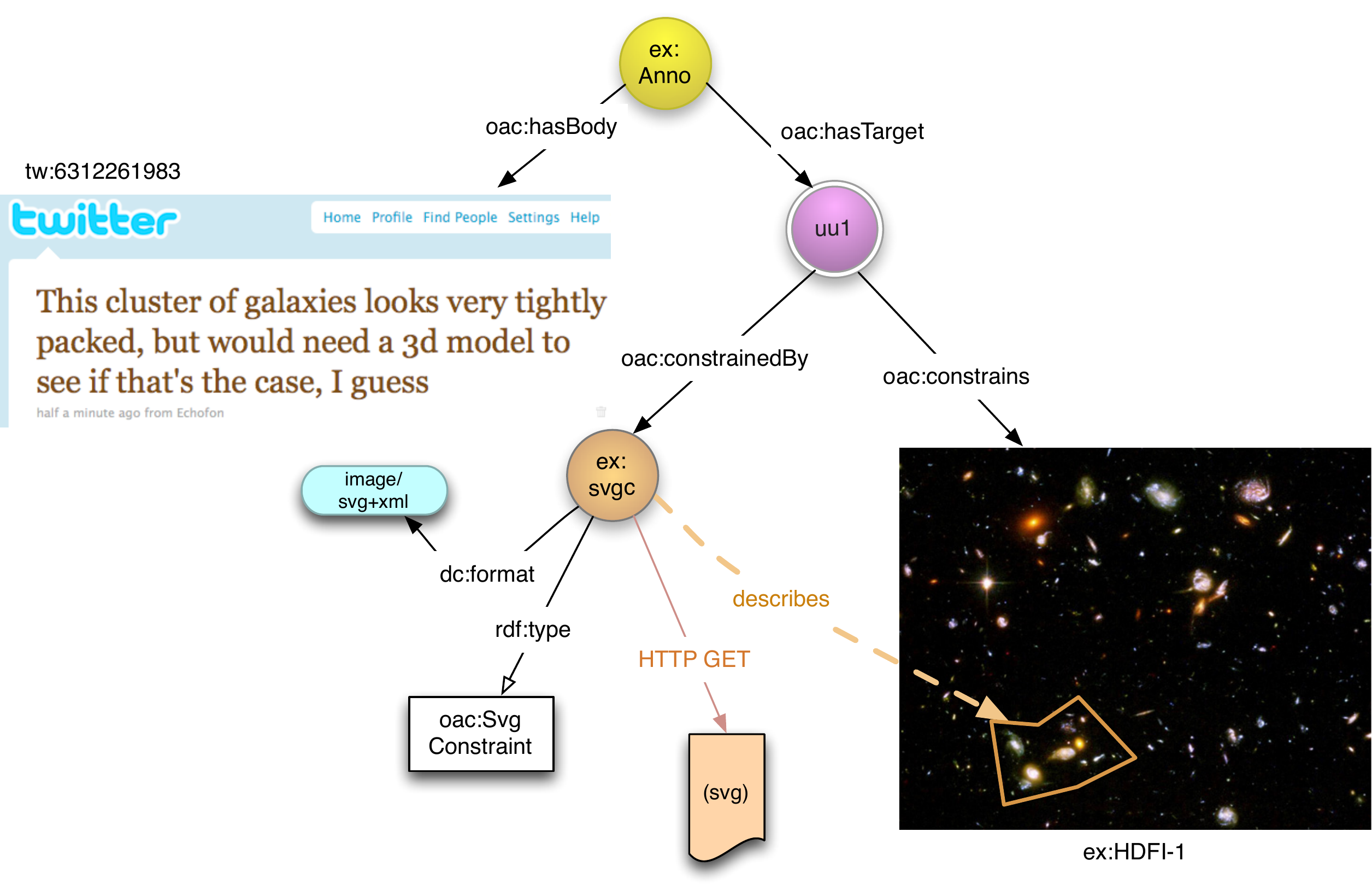}
  \caption{Annotating media segments using an SVG constraint.}
  \label{fig:oac_model_svg_constraint_example}
\end{figure}

The nature of the constraint description will be dependent on the type of the resource for which the segment is being conveyed. It is then up to the annotation client to interpret the segment description with respect to the full resource. Figure~\ref{fig:oac_model_svg_constraint_example} shows an example in which an area within an image is described by an SVG path element. The document containing the SVG specification is identified by a dereferencable HTTP URI and a specialized type \texttt{oac:SvgConstraint} in combination with a \texttt{dc:format} property
informs the client about the type of constraint it needs to deal with.

Alternatively, it is also possible to include the constraint information inline within the annotation document using the same technique as used for including the body. The \texttt{oac:Constraint} is given a URN (normally a urn:UUID) and then the constraint information is included as the value of the \texttt{cnt:chars} property. The requirements for doing this are the same as for including the \texttt{oac:Body} inline within the annotation document. For more complex use cases it also possible to express constraints in RDF and to apply constraints also on the body of an annotation.

One goal of the ongoing Open Annotation demonstrator activities is to collect real-world constraint definitions from various use cases and to specify them in the context of the OA model. We hope that this also serves as as feedback loop for possible enhancements or additional URI fragment specifications.


\section{Robust Annotations over Time}\label{sec:robust_annotations}

It must be stressed that different agents may create the \emph{annotation}, \emph{body} and \emph{target} at different times. For example, Alice might create an annotation saying that Bob's YouTube video annotates Carol's Flickr photo. Also, being regular Web resources, the body and target are likely to have different representations over time. Some annotations may apply irrespective of representation, while others may pertain to specific representations. In order to provide the ability to accurately interpret annotations past their publication, the Open Annotation Model introduces three ways to express temporal context. The manner in which these three types of annotations use the \texttt{oac:when} property, which has a datetime as its value, distinguishes them.

A \emph{Timeless Annotation} applies irrespective of the evolving representations of body and target; it can be considered as if the annotation references the semantics of the resources. For example, an annotation with a body that says ``This is the front page of CNN'' remains accurate as representations of the target \url{http://cnn.com/} change over time. Timeless annotations don’t make use of the \texttt{oac:when} property.

\begin{figure}
  \centering
  \includegraphics[width=0.75\textwidth]{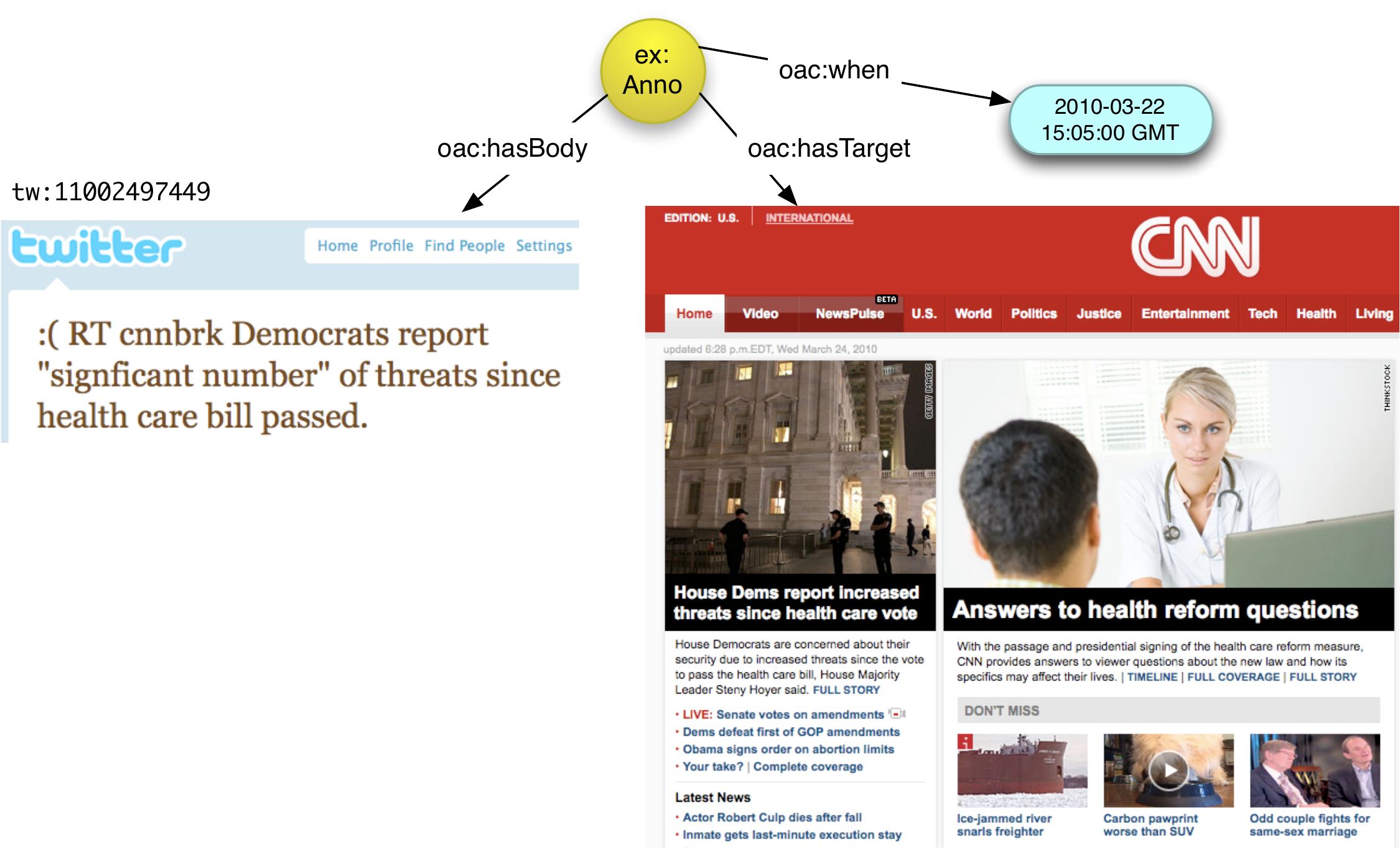}
  \caption{A Uniform Time Annotation example.}
  \label{fig:uniform_time_example}
\end{figure}

A \emph{Uniform Time Annotation} has a single point in time at which all the resources involved in the annotation should be considered. This type of annotation has the \texttt{oac:when} property attached to the \texttt{oac:Annotation}. For example, if Alice recurrently publishes a tweet that comments on a story on the live CNN home page, an annotation that has the cartoon as body and the CNN home page as target would need to be handled as a Uniform Time Annotation in order to provide the ability to match up correct representations of body and target. Figure~\ref{fig:uniform_time_example} shows how Uniform Time Annotations can be represented using the Open Annotation Model.

A \emph{Varied Time Annotation} has a body and target that need to be considered at different moments in time. This type of Annotation uses the \texttt{oac:when} property attached to an \texttt{oac:WebTimeConstraint} node, which is a specialization of \texttt{oac:Constraint}, for both body and target. If, in the aforementioned example, Alice would have the habit to publish a cartoon at \url{http://example.org/cartoon} when the mocked article is no longer on the home page, but still use \url{http://cnn.com} as the target of her annotation, the Varied Time Annotation approach would have to be used.

This temporal information can be used to recreate the annotation as it was intended by reconstructing it with the time-appropriate body and target(s). Previous versions of Web resources exist in archives such as the Internet Archive, or within content management systems such as MediaWiki's article history, however they are divorced from their original URI. Memento~\cite{Van-de-Sompel:2009fk,Sompel:2010fk}, which is a framework that proposes a simple extension of HTTP in order to connect the original and archived resources, can be applied for recreating annotations. It leverages existing HTTP capabilities in order to support accessing resource versions through the use of the URI of a resource and a datetime as the indicator of the required version. In the framework, a server that host versions of a given resource exposes a TimeGate, which acts as a gateway to the past for a given Web resource. In order to facilitate access to a version of that resource, the TimeGate supports HTTP content negotiation in the datetime dimension. Several mechanisms support discovery of TimeGates, including HTTP links that point from a resource to its TimeGate(s)~\cite{Sanderson:2010fk}.


\section{Related Work}\label{sec:related_work}

In this section we give an overview of existing work in the area of Web annotations. After
summarizing general works about annotations and annotation interoperability, we analyze the features
of existing Web annotation models and compare them with those of the Open Annotation Model.

\subsection{Annotations and annotation interoperability}

Annotations have a long research history, and unsurprisingly the research perspectives
and interpretations of what an \emph{annotation} is supposed to be vary widely. Agosti
et al.~\cite{Agosti:2007uq} provide a comprehensive study on the contours and complexity
of annotations. A representative discussion on how annotations can be used in various
scholarly disciplines is given by Bradley~\cite{Bradley:2008kx}. He describes how annotations
can support interpretation development by collecting notes, classifying resources, and
identifying novel relationships between resources.

The different forms and functions that annotations can take are analyzed by
Marshall~\cite{Marshall:2000kx}. She distinguishes between \emph{formal} and \emph{informal} annotations, 
whereby formal annotations follow structural standards and informal ones are unstructured. 
Furthermore, Marshall divides into \emph{implicit} annotations that are intended for sharing and \emph{explicit} annotations of personal nature, often interpretable only by the original creator.
Further divisions defined by Marshall with regard to the function of an
annotation include \emph{permanent} vs. \emph{transient}, \emph{annotation as
writing} vs. \emph{annotation as reading}, \emph{extensive} vs.
\emph{intensive}, \emph{published} vs. \emph{private} and \emph{institutional}
vs. \emph{workgroup} vs. \emph{individual}. The difference between personal and public annotations in a digital environment is further investigated in a study by Marshall and Brush~\cite{Marshall2004}. They derive design implications for
annotation systems, e.g. regarding find and filtering requirements, and user
interface strategies for processing and sharing annotations.

A taxonomy of
annotation types and marking symbols used by readers of scholarly documents is
presented by Qayyum~\cite{Qayyum2008}. His taxonomy is derived from the results
of a user study conducted with students reading research articles in a private
as well as a collaborative digital setting. A related recent effort is presented
by Blustein et al.~\cite{Blustein2011}. In their field study, conducted over the
course of three years, they identify six purposes for scholarly annotation:
\emph{interpretation}, \emph{problem-working}, \emph{tracing progress},
\emph{procedural annotations}, \emph{place marking and aiding memory} and
\emph{incidental markings}.

\subsection{Web annotation models}

The idea of publishing user annotations on the Web is not new. Annotea~\cite{Kahan:2001vn} was specified
more than a decade ago and defines a data model and protocol for uploading, downloading, and modifying annotations. Since
the Web has changed over time and now also comprises non-document Web resources, the Annotea
model soon became insufficient for many annotation use cases, as we explained at the beginning of this article. Annotea
extensions, such as Co-Annotea~\cite{Hunter:2008ab} or LEMO~\cite{Haslhofer:2009ve}, were developed to deal with
the Annotea shortcomings and to take into account emerging architectural styles, such as RESTful Web Services, or Linked Open Data. Recent Web annotation model specification efforts include the M3O ontology~\cite{Saathoff:2010vn}, the Annotation Ontology~\cite{Ciccarese2011}, and the Open Annotation Model, which we presented in this article. 

\begin{figure}
  \centering
  \includegraphics[width=1\textwidth]{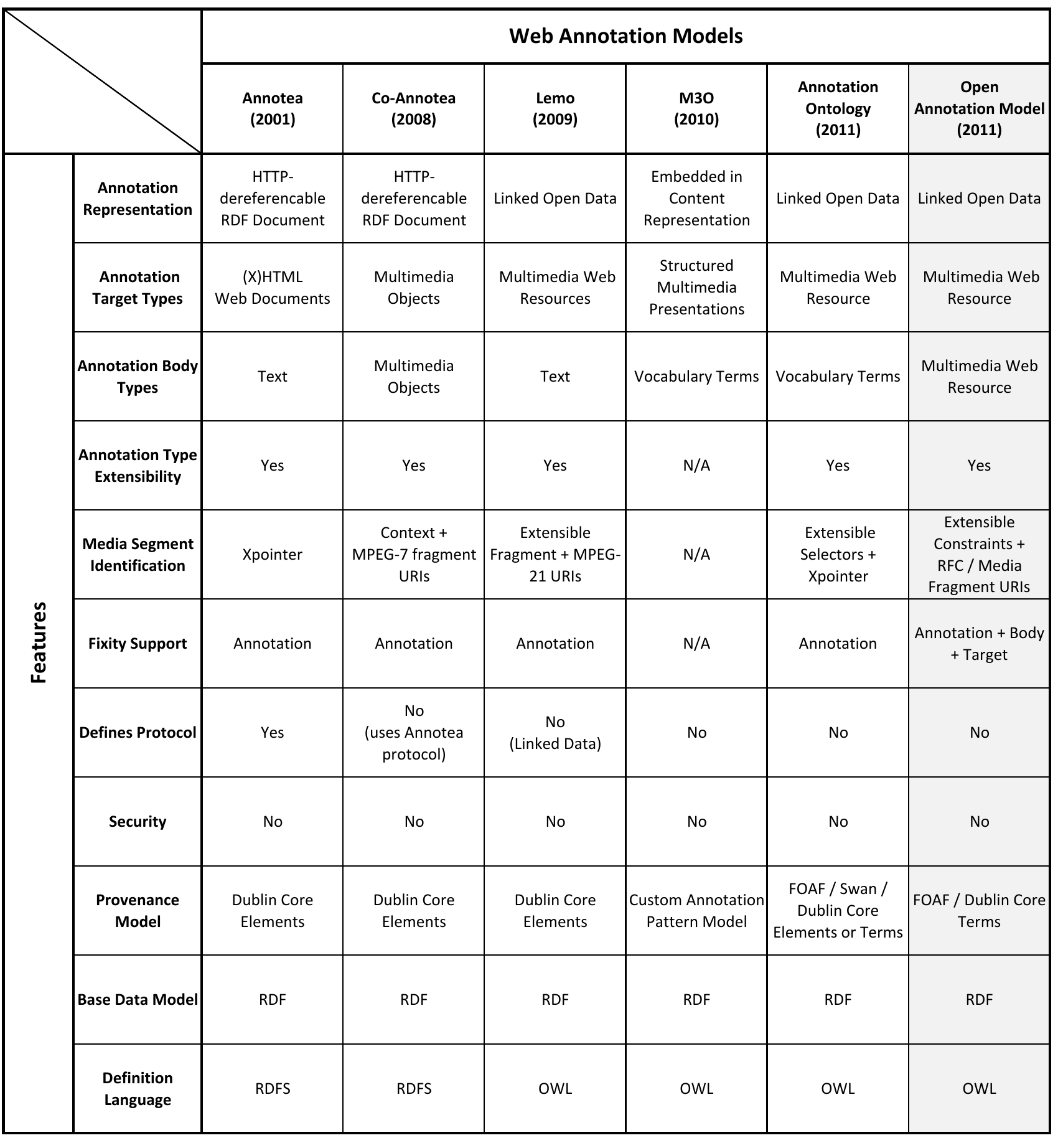}
  \caption{Feature analysis of existing Web annotation models.}
  \label{fig:existing_ann_models_ft_analysis}
\end{figure}

In Figure~\ref{fig:existing_ann_models_ft_analysis} we present the results of a feature comparison we performed
across the previously mentioned models. The models are timely ordered by their publication year, and the feature selection is based on the requirements and use case descriptions we found in the model documentations. Although we cannot generalize from this representative
set of annotation models and features, we can observe hat annotation models have continually been adapted to emerging standards, needs, and architectures: Linked Open Data is increasingly adopted for publishing and sharing annotations, multimedia resources
can now be annotated also by other multimedia resources, extensibility has become a key requirement, security and provenance are being outsourced to other models, and standard segment identification mechanisms are being combined with custom solutions (context, fragment, selectors) to capture complex domain-specific needs.
 
The Annotation Ontology, which is an open ontology for the annotation of scientific documents on the Web, is technically very similar to the Open Annotation Model. However, they differ e.g., in terms of how fragments are being expressed: by representing constraints and constraint targets as first-class resources the Open Annotation Model supports direct addressing of fragments, thus enabling use cases where different users annotate the same fragment, or search scenarios where annotations are retrieved by fragment. Furthermore, the Open Annotation Model supports structured annotation bodies and allows to overlay semantic statements pertaining to one or more annotation targets, which offers potentially more flexibility, e.g., for use cases of entity and entity-relation extraction in scientific literature.

A related strand of research concerns models for \emph{social tagging} of Web resources. Hunter~\cite{Hunter2009} describes tags as ``\emph{a subclass of annotations that comprise simple, unstructured labels or keywords assigned to digital resources to describe and classify the digital resource}''. A comparison of tagging ontologies is presented by Kim et al.~\cite{KimEtAl2008}. They
survey the state of the art in tagging models and identify three building blocks common to existing tagging models: \emph{taggers}, the \emph{tags} themselves and the \emph{resources} being tagged.

Semantic annotations features can also be found in multimedia metadata frameworks such as MPEG-7 and multimedia metadata ontologies such as COMM~\cite{Arndt:2007uq} or the recent \emph{W3C Ontology for Media Resources}\footnote{\url{http://www.w3.org/TR/mediaont-10/#example3}} specification, which provides a core metadata vocabulary for media resources on the Web. It defines two metadata properties that can be used for the textual description of a media resource (fragment) or for relating RDF files or named graphs to a media resource. Other ontologies were designed to embed annotations directly into the multimedia content representation. The M3O Ontology~\cite{Saathoff:2010vn}, for instance, allows the integration of annotations with SMIL and SVG documents. On the contrary, the Open Annotation Model is more in line with the previously discussed Web Annotation models. It treats annotations as first class Web resources, which can exist independently from the content or metadata representations of media objects. This design choice is motivated by a set of scholarly use cases, which require that multimedia content objects can be annotated any time after the content production and metadata extraction process.

\subsection{Media segment identification}\label{subsec:rel_work_media_segments}

Early related work on the issue of describing segments in multimedia resources can be traced back to research on linking in hypermedia documents (cf.~\cite{Hardman:1994zr}). For describing segments using a non-URI based mechanism one can use MPEG-7 Shape Descriptors (cf.~\cite{Nack:1999ly}) or terms defined in a dedicated multimedia ontology. SVG~\cite{svg:2003bh} and MPEG-21~\cite{ISO/IEC:2006qf} introduced XPointer-based URI fragment definitions for linking to segments in multimedia resources. The Temporal URI specification~\cite{Pfe07} addresses a temporal segment in a time-based media resource through a defined URL query parameter ('t='). YouTube supports similar direct linking to a particular point in time in a video
using a fragment URI. The Media Fragments URI Specification~\cite{VanDeursen:2010} is a W3C Working Draft that introduces a standard, URI-based approach for addressing temporal, spatial and track sub-parts of any non-textual media content, thus making audiovisual media segments first class citizens on the Web~\cite{Hausenblas:LDOW09}.

\subsection{Robustness of Web resources}

The ephemeral nature of Web resources and methods to deal with that problem have been studied from the early years of the Web on. Phelps and Wilensky~\cite{Phelps:2000ys} proposed to decorate hyperlinks with lexical signatures to re-find disappeared web resources. Recent works include Klein et al.~\cite{Klein:2011kx}, who proposed to compute lexical signatures from the link neighborhood of a Web page, and Morishima et al.~\cite{Morishima:2009vn} who describe a method to fix broken links when link targets have moved. The problem has also been realized in the Linked Data context and solutions like DSNotify~\cite{Popitsch:2011zr} were proposed to re-find resources by their representations.


\section{Summary and Future Directions}\label{sec:summary}

We apply a generic and Web-centric conception to the various facets annotations can have and regard an annotation as association created between one \emph{body} resource and other \emph{target} resources, where the body must be somehow \emph{about} the target. This conception lead to the specification of the Open Annotation Model, which originates from activities in the Open Annotation Collaboration and aims at building an interoperable environment for publishing annotations on the Web.

At the time of this writing, the Open Annotation Model is still in beta stage and is currently implemented in several demonstration projects covering use cases ranging from annotating medieval manuscripts, over annotating online maps, to annotating online video segments. As part of these demonstration projects client APIs are being developed for various programming environments to ease model adoption for developers. We expect to obtain user and developer feedback from these projects, which should further refine the Open Annotation Model.

Pursuing better integration of the proposed segment identification approach with the W3C Media Fragment URI specification is on our research agenda. However, this requires an extension mechanism in that specification, which is not within the scope of the responsible working group at the moment. Also the question of how to model multiple annotation targets and how to interpret this information correctly, is currently being discussed. Finally, as a first outcome of the demonstration projects, we observed that a common set of use cases is annotating data with other data, rather than with information intended for human consumption. This raises the question of how to model \emph{Data Annotations} in an interoperable way.

As a final result, we expect a data model, which provides an interoperable method of expressing annotations such that they can easily be shared between platforms, with sufficient richness of expression to satisfy also complex annotation scenarios.


\begin{acknowledgements}

The work has partly been supported by the European Commission as part of the eContentplus program (EuropeanaConnect) and by a Marie Curie International Outgoing Fellowship within the 7th Europeana Community Framework Program. The development of OAC is funded by the Andrew W. Mellon foundation.

\end{acknowledgements}


\end{document}